\documentclass{article}
\usepackage{spconf,amsmath,graphicx}
\usepackage{cite}
\usepackage{hhline}
\usepackage[table]{xcolor}
\usepackage{comment,threeparttable,multirow,array}
\usepackage{subfig}
\usepackage{enumitem}
\usepackage{color,soul}

\title{Privacy-Preserving Deep Neural Networks with Pixel-based Image Encryption Considering Data Augmentation in the Encrypted Domain}
\name{Warit Sirichotedumrong, Takahiro Maekawa, Yuma Kinoshita and Hitoshi Kiya\thanks{This work was partially supported by Grant-in-Aid for Scientific
Research(B), No.17H03267, from the Japan Society for the Promotion Science.}}
\address{Tokyo Metropolitan University, Asahigaoka, Hino-shi, Tokyo, 191-0065,
Japan}
\makeatletter
\newcommand{\vect}[1]{\boldsymbol{#1}}

\makeatother
\begin{document}
\ninept

\maketitle

\begin{abstract}
We present a novel privacy-preserving scheme for deep neural networks
(DNNs) that enables us not to only apply images without visual
information to DNNs for both training and testing but to also consider
 data augmentation in the encrypted domain for the first time. In
this paper, a novel
pixel-based image encryption method is first proposed for privacy-preserving
DNNs. In addition, a novel adaptation network is considered that
reduces the influence of image encryption. In an experiment, the
proposed method is applied to a well-known network, ResNet-18, for
image classification. The experimental results demonstrate that conventional
privacy-preserving machine learning methods including the
state-of-the-arts cannot be applied to data augmentation in the
encrypted domain and that the proposed method outperforms them in terms of classification accuracy.
\end{abstract}
\begin{keywords}
Deep learning, deep neural network, image encryption, privacy-preserving
\end{keywords}
\section{Introduction}
\label{sec:intro}
The spread of deep neural networks (DNNs) has greatly contributed
to solving complex tasks for many applications\cite{Donahue2014,Krizhevsky2012},
such as for computer vision, biomedical systems, and information technology.
Deep learning utilizes a large amount of data to extract 
representations of relevant features, so the performance is significantly improved\cite{Tishby2015,michael2018on}.
However, there are security issues when using deep learning in cloud environments to train and test data, 
such as data privacy, data leakage, and unauthorized data
access. Therefore, privacy-preserving DNNs have become an urgent challenge.

In this paper, we propose a novel privacy-preserving method for DNNs
that enables us to not only apply images without visual information to
DNNs for both training and testing but to also carry out data
augmentation in the encrypted domain for the first time. Data
augmentation, which is a technique for creating new training data from
existing data, is widely used in DNN-based methods because it is easy to implement and 
is effective. Augmented data can be utilized to
improve model robustness and increase accuracy. Since augmentation
produces a huge amount of data, it is required that it has to be performed on a
cloud server in order to reduce the amount of data traffic. This paper
is among the first to discuss data augmentation in the encrypted
domain.

Various methods have been proposed for privacy-preserving computation. The methods are classified into two types: perceptual
encryption-based\cite{apsipa_svm, tanaka,Ito_2009,Tang_2014,KURIHARA2015encryption,Kuri_2017,CHUMAN2017ICASSP,CHUMAN2017IEICE,ChumanIEEETrans,sirichotedumrong_kiya_2019,crypto_compress,2016_Gaata} and homomorphic encryption (HE)-based
\cite{Araki2016,Araki2017,Lu2016UsingFH,Aono2015,Shokri2015,Phong2018,cryptonets,8350963}.
As described in Section 2, HE-based methods are the most secure options for
privacy preserving computation,
but they are applied to only limited DNNs\cite{Shokri2015,Phong2018,cryptonets,8350963}.
Therefore, the HE-based type does not support state-of-the-art DNNs yet.
Moreover, data augmentation has to be done before encryption.
In contrast, perceptual encryption-based methods
have been seeking a trade-off in security to enable other requirements, such
as a low processing demand, bitstream compliance, and signal processing
in the encrypted domain\cite{apsipa_svm, tanaka,Ito_2009,Tang_2014,KURIHARA2015encryption,Kuri_2017,CHUMAN2017ICASSP,CHUMAN2017IEICE,sirichotedumrong_kiya_2019,ChumanIEEETrans,crypto_compress,2016_Gaata}.
A few methods were applied to machine learning algorithms in previous works\cite{apsipa_svm, tanaka}.
The first encryption method\cite{KURIHARA2015encryption,Kuri_2017,CHUMAN2017ICASSP,CHUMAN2017IEICE,sirichotedumrong_kiya_2019,ChumanIEEETrans}
to be proposed for encryption-then-compression (EtC)
systems, was demonstrated to be applicable to traditional machine
learning algorithms, such as support vector machine (SVM)\cite{apsipa_svm}.
However, the block-based encryption method has never been applied to DNNs.
Another method\cite{tanaka} was applied to image
classification with DNNs, in which an adaption network is added prior
to DNNs to avoid the influence of image encryption.
However, the accuracy of image classification is lower than that of using plain
images, and, moreover, data augmentation in the encrypted domain cannot be applied to the conventional method.

In an experiment, we compare the proposed method with conventional perceptual encryption-based methods.
The experimental results show that
the proposed methods with DNNs performs better in classification than conventional block-based and pixel-based encryption schemes.
In addition,
it is proved that the proposed encryption allows us to perform
data augmentation in the encrypted domain.
\section{Related Works}
\label{sec:preparation}

\subsection{Visual Information Protection}
\label{ssec:visual}

Security mostly refers to protection from adversarial
forces. This paper focuses on protecting visual information
that allows us to identify an individual, the time, and
the location of the taken photograph. Untrusted platforms
and unauthorized users are assumed to be adversaries.

Various perceptual image encryption methods\cite{apsipa_svm, tanaka,Ito_2009,Tang_2014,KURIHARA2015encryption,Kuri_2017,sirichotedumrong_kiya_2019,ChumanIEEETrans,CHUMAN2017ICASSP,CHUMAN2017IEICE,crypto_compress,2016_Gaata}
have been proposed for protecting the visual information of images. Compared with full
encryption with provable security like homomorphic encryption (HE), they generally have a low
computational cost and can offer encrypted data robust against
various kinds of noise and errors. In addition, some of them aim to consider
both security and efficient compression so that they can be adapted to cloud
storage and network sharing \cite{KURIHARA2015encryption,Kuri_2017,sirichotedumrong_kiya_2019,ChumanIEEETrans,CHUMAN2017ICASSP,CHUMAN2017IEICE,crypto_compress}.  
However, with the exception of a few previous pieces of work, most conventional perceptual encryption methods have never been
considered for application to machine learning algorithms\cite{apsipa_svm, tanaka}.
\begin{figure}[!t]
\captionsetup[subfigure]{justification=centering}
\centering
\subfloat[Original image ($X \times Y$ = $96 \times 96$)]{\includegraphics[clip, height=3cm]{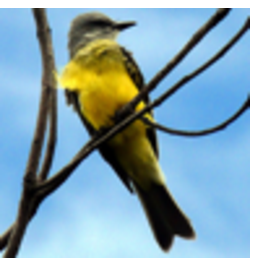}
\label{fig:label-A}}
\hfil
\subfloat[Block-based encryption\cite{KURIHARA2015encryption,Kuri_2017} (Block size=$4 \times 4$)]{\includegraphics[clip, height=3cm]{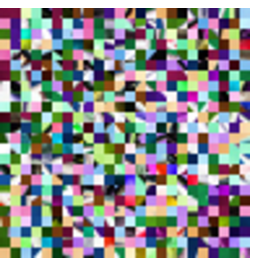}
\label{fig:label-C}}
\\
\subfloat[Block-based encryption\cite{tanaka} (Block size=$4 \times 4$)]{\includegraphics[clip, height=3cm]{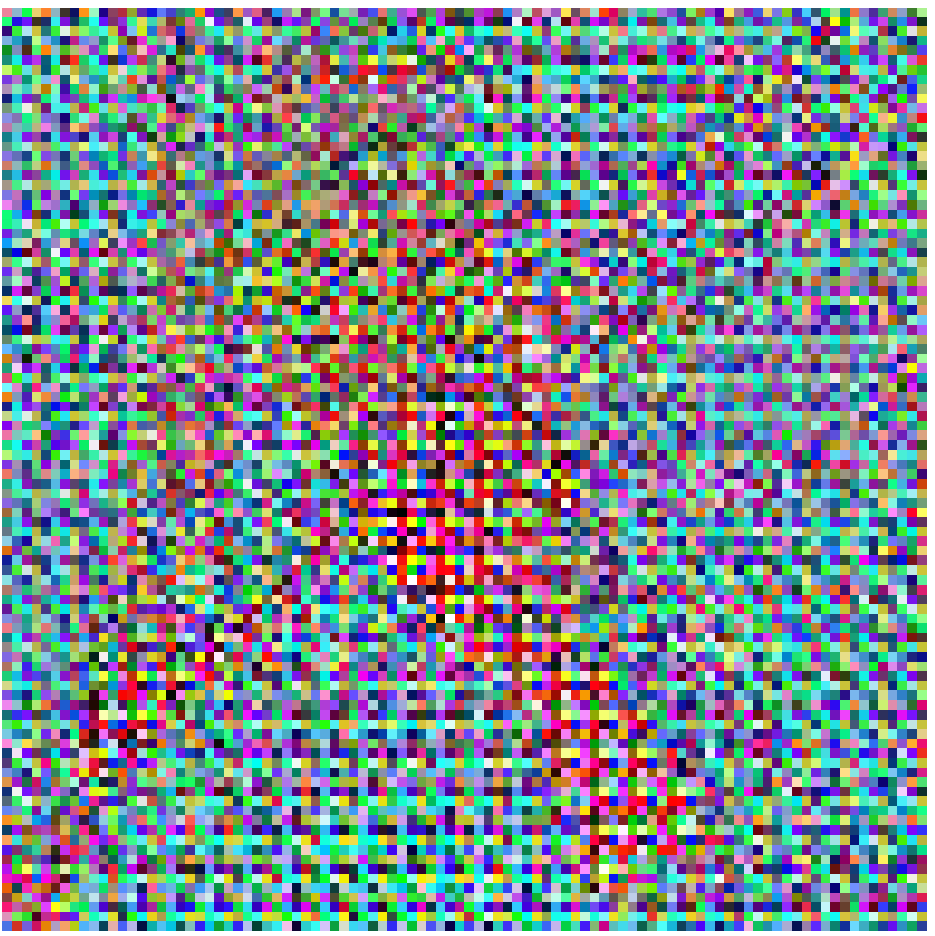}
\label{fig:label-C}}
\hfil
\subfloat[Pixel-based image encryption\cite{2016_Gaata}]{\includegraphics[clip, height=3cm]{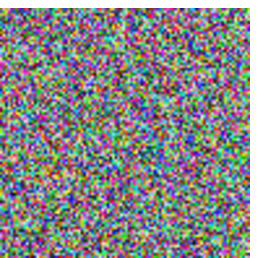}
\label{fig:label-D}}
\caption{Examples of images encrypted by conventional schemes}
\label{fig:eximages}

\end{figure}
\subsection{Privacy-Preserving Machine Learning}
\label{ssec:privacy}

As mentioned above, two perceptual image encryption methods have been studied for
privacy-preserving machine learning so far. 
The first\cite{apsipa_svm, tanaka,Ito_2009,Tang_2014,KURIHARA2015encryption,Kuri_2017,sirichotedumrong_kiya_2019,ChumanIEEETrans,CHUMAN2017ICASSP,CHUMAN2017IEICE,crypto_compress}
is applicable to tradition machine learning algorithms, such as support vector machine (SVM), k-nearest neighbors (KNN), and random forest even under the use of the kernel trick \cite{apsipa_svm}.
However, its block-based encryption method has never been applied to DNNs. The other\cite{tanaka} was applied to image classification with
DNNs, but the accuracy is lower than that when using plain images, and the influence of data augmentation in the encrypted domain cannot be avoided yet. Examples of encrypted images are shown in Fig.\,\ref{fig:eximages}.

Alternatively, privacy-preserving machine learning methods with homomorphic encryption (HE) \cite{Shokri2015,Phong2018,cryptonets,8350963} have been studied. One is
CryptoNet\cite{cryptonets}, which can apply HE to the influence stage of CNNs.
CryptoNet has very high computational complexity, so a dedicated low
computer convolution core architecture for CryptoNet was proposed and implemented with a CMOS technology\cite{8350963}. 
In CryptoNet, all activation functions and the loss function must be polynomial functions. Therefore, 
it cannot be applied to state-of-the-art DNNs. Moreover, CryptoNet
does not allow us to carry out data augmentation in the encrypted
domain, in addition to the high computation complexity.

One approach with HE has been proposed for privacy-preserving
weight transmission for multiple owners who wish to apply a machine
learning method over combined data sets\cite{Shokri2015,Phong2018}.
However, this approach can not be applied to network training in the
encrypted domain.

In this paper, we aim to consider a method that enables not only
network training in the encrypted domain but also data augmentation
in the encrypted domain. This paper is among the first to discuss data augmentation in the encrypted domain.

\section{Proposed methods}
\label{sec:proposed}
\subsection{Overview of Privacy Preserving DNNs}
\label{ssec:overview}

\begin{figure}[t]
\centering
\includegraphics[width =8.5cm]{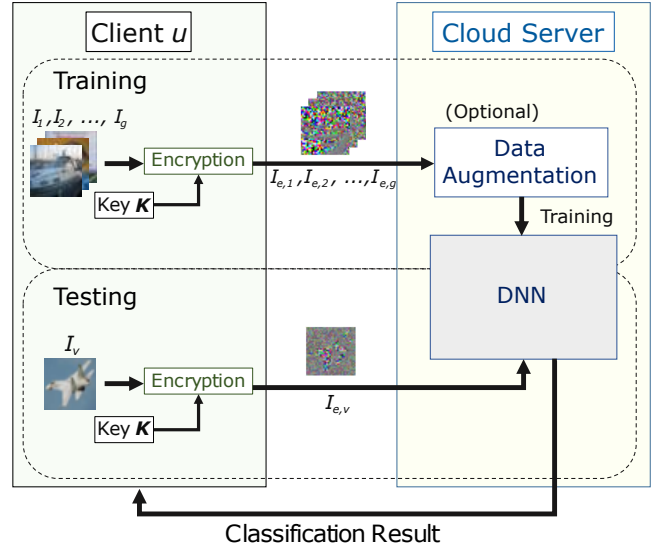}
\caption{Scenario}
\label{fig:scenario}
\end{figure}

Figure\,\ref{fig:scenario} illustrates the scenario used in this paper. In the training process, a client $u$ encrypts all training images ($I_1, I_2,\ldots, I_g$) to protect the
visual information of the training images by using a secret key set, $\vect{K}$, and sends the encrypted images ($I_{e,1}, I_{e,2},\ldots, I_{e,g}$) to a cloud server.

In the testing process, the client $u$ encrypts a testing image ($I_v$) by using a secret key set, $\vect{K}$, and sends the encrypted image $I_{e,v}$ to a server.
The server solves a classification problem with an image classification model trained in advance, and then returns the classification results to the client.

Note that the server has no secret key, so clients are able to control the privacy of images by themselves even when the classification process is done
in the server. As shown in Fig.\,\ref{fig:scenario}, data augmentation is carried out in both the client and the server, or by the server.
In other words, data augmentation can be done after encryption.
 However, for all conventional encryption methods, data augmentation has to be done before encryption. 

\subsection{Proposed Image Encryption}
\label{ssec:proposed_enc}
 \begin{figure}[t]
\centering
\includegraphics[width =8.5cm]{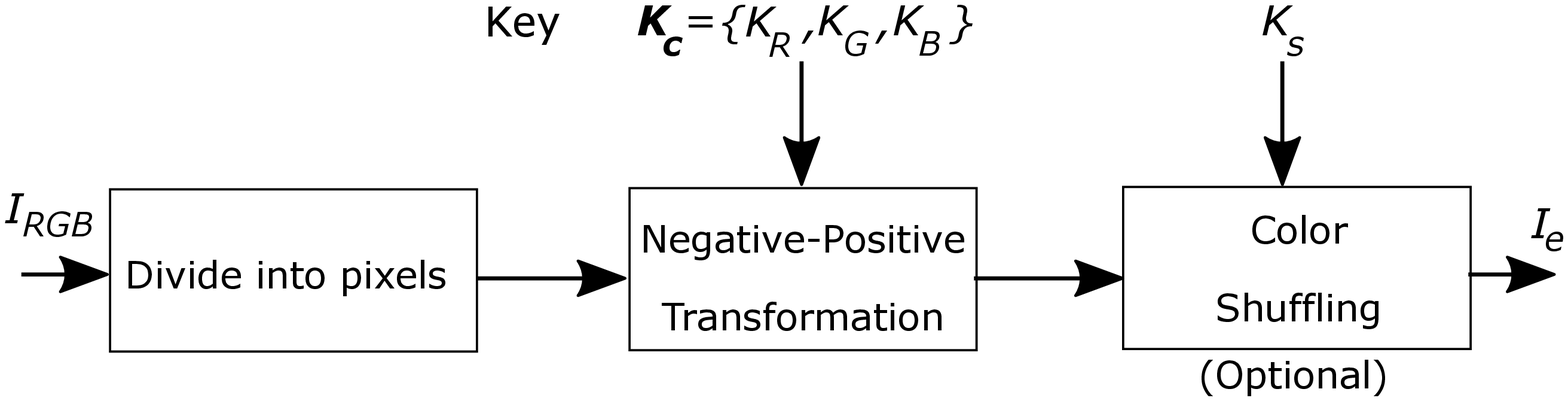}
\caption{Proposed image encryption}
\label{fig:enc_steps}
\end{figure}
We present a novel perceptual image encryption method that aims not to only relax
 the limitations of using encrypted images in DNNs but to also use data augmentation in the encrypted domain.
 In the block-based encryption\cite{tanaka}, the number of feature maps has to be reduced due to block adaptation.
In contrast, the proposed encryption is a pixel-based encryption method that enables an adaptation network with a small number of parameters, so the resolution of an encrypted image can be preserved.
 Moreover, the proposed encryption method can provide data robust against ciphertext-only attacks compared with the conventional one\cite{tanaka}.
 
To generate an encrypted image ($I_e$) from a color image, $I_{RGB}$, the following steps are carried out, as shown in Fig.\,\ref{fig:enc_steps}. Note that the color shuffling (Step 3) is an optional encryption step to enhance security.
\begin{itemize}[nosep]
  \item [1)] Divide $I_{RGB}$ with $X \times Y$ pixels into pixels.
  \item [2)] Individually apply negative-positive transformation to each pixel of each color channel, $I_R$, $I_G$, and $I_B$, by using a random binary
   integer generated by secret keys $\vect{K_c}=\{K_R, K_G, K_B\}$. In this step, a transformed pixel value of
  the $i$-th pixel, $p'$, is calculated using 

	\begin{equation}
	\label{eq:p_nega-pos}
	p'=
	\left\{
	\begin{array}{ll}
	p & (r(i)=0) \\
	p \oplus (2^L-1) & (r(i)=1)
	\end{array} ,
	\right.
	\end{equation}
	where $r(i)$ is a random binary integer generated by $K_c$. $p$ is
	the pixel value of the original image with $L$ bit per pixel. The
value of the occurrence probability $P(r(i))=0.5$ is used to invert bits
randomly\cite{ChumanIEEETrans}.

  \item [3)] (Optional) Shuffle three color components of each pixel by using an integer randomly selected from six integers 
  generated by a key $K_s$ as shown in Table\,\ref{tbl:color_shuf}.
\end{itemize}

\begin{table}
\centering
\small
\caption{Permutation of color components for random
integer. For example, if random integer is
equal to 2, red component is replaced by green
one, and green component is replaced by red one while blue component is not
replaced.}
\label{tbl:color_shuf}
\begin{tabular}{|>{\centering\arraybackslash}m{1in}||>{\centering\arraybackslash}m{1cm}|>{\centering\arraybackslash}m{1cm}|>{\centering\arraybackslash}m{1cm}|}
\hline
\multirow{2}{*}{Random Integer} & \multicolumn{3}{c|}{Three Color Channels
}\\
\hhline{~---}
& R & G & B\\
\hline
0 & R & G & B\\
\hline
1 & R & B & G\\
\hline
2 & G & R & B\\
\hline
3 & G & B & R\\
\hline
4 & B & R & G\\
\hline
5 & B & G & R\\
\hline
\end{tabular}
\end{table}

\begin{figure}[!t]
\captionsetup[subfigure]{justification=centering}
\centering
\subfloat[Negative-positive transformation]{\includegraphics[clip, height=3cm]{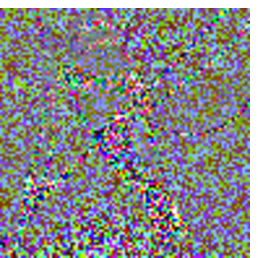}
\label{fig:label-A}}
\hfil
\subfloat[Negative-positive transformation and color shuffling]{\includegraphics[clip, height=3cm]{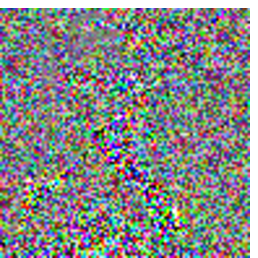}
\label{fig:label-B}}
\caption{Examples of images encrypted by proposed method}
\label{fig:proposed_images}
\vspace{-0.5cm}
\end{figure}

Examples of images encrypted by using the proposed scheme are shown in Fig.\,\ref{fig:proposed_images}, where Fig.\,\ref{fig:eximages}(a) is the original one.
It is proved that the visual information of images was protected as in Fig.\,\ref{fig:eximages}.

\subsection{Data Augmentation}
\label{ssec:aumentation}
To solve complex tasks,
a large amount of data is necessary to train DNNs.
Data augmentation aims to enlarge the number of data points used for training
and enables us to avoid the overfitting of DNNs.
Many data augmentation techniques have already been proposed,
e.g., horizontal/vertical flip, random crop, random rotation, cutout,
 and random erasing\cite{rand_erasing}.
Although data augmentation is very useful for increasing the performance of DNNs,
there are no image encryption methods that consider data augmentation.
For this reason, data augmentation has to be carried out in each client before encryption when privacy-preserving DNNs are employed.

In this paper, we consider performing augmentation in a cloud server,
namely, images are augmented after encryption.
Here, the following well-known techniques are utilized for data augmentation:
\begin{itemize}[nosep]
  \item \textit{Horizontal/vertical} flip:
    flips original images horizontally or vertically.
  \item \textit{Shifting:}
    shifts pixel locations of original images
    on both horizontal and vertical axes by number of pixels.
\end{itemize}

In Section\,\ref{sec:experiments},
the effect of data augmentation in a cloud server is experimentally analyzed.
The results show that the block-based encryption method\cite{tanaka} was heavily affected by carrying out the augmentation in the encrypted domain.
In contrast, the proposed encryption maintained a high classification performance, even when the augmentation was carried out after encryption.

\subsection{Security Evaluation}
\label{ssec:security}

Security mostly refers to protection from adversarial forces. Various attacking strategies, such as the known-plaintext attack (KPA) and chosen-plaintext attack (CPA), should be considered\cite{ChumanIEEETrans,CHUMAN2017ICASSP,CHUMAN2017IEICE}.
Here, we consider brute-force attacks as ciphertext-only attacks.

If an image with $X \times Y$ pixels is divided into pixels, the number of pixels $n$ is given by
\begin{equation}
\label{eq:blocknum}
n = X \times Y.
\end{equation}

The key spaces of negative-positive transformation ($N_{np}$) and color component shuffling ($N_{col}$) are represented by
\begin{equation}
{N_{np}(n)} = 2^{3n}, {N_{col}(n)} = \bigl( { }_3 P { }_3 \bigr)^{n} = 6^{n}.
\end{equation}

Consequently, the key space of images encrypted by using the proposed encryption scheme, $N(n)$, is represented by the following.
\begin{equation}
\begin{array}{ll}
N(n) & = {N_{np}(n)} \cdot {N_{col}(n)}
\\
& =2^{3n} \cdot 6^{n}
\end{array}
\end{equation}

In contrast, in Tanaka's method\cite{tanaka}, $I_{RGB}$ with $X \times Y$ pixels is divided into blocks each with $4 \times 4$ pixels, and 
each block is split into upper 4-bit and lower 4-bit images to generate 6-channel image blocks. Then, the intensities of randomly selected pixels are
reversed. Eventually, the pixels in each block are shuffled with the same pattern.

The key space of Tanaka's method\cite{tanaka}, $N_{tanaka}$, is given by
\begin{equation}
\label{eq:blocknum}
N_{tanaka}=96!\cdot 2^{96}.
\end{equation}

Therefore, since $N(n)>>N_{tanaka}$, the proposed encryption has a larger key space than Tanaka's method.
 
\subsection{Adaptation Network}
\label{ssec:adaptation}

We propose a novel adaptation network for DNNs that aims to adapt images encrypted by the proposed encryption method to make the images compatible with
DNNs. Since the proposed encryption method is a pixel-based one, the proposed adaptation network consists of simple $1 \times 1$-convolutional layers.

\begin{figure}[t]
\centering
\includegraphics[width =8.5 cm]{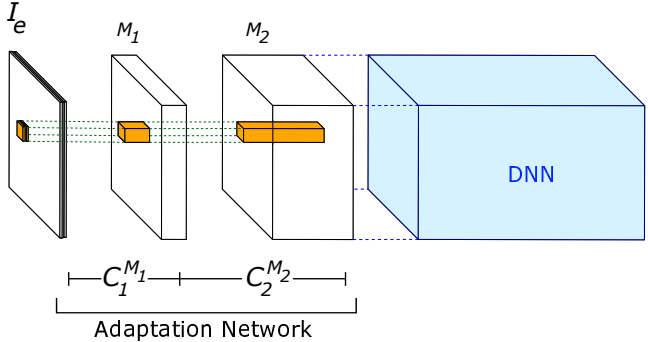}
\caption{Proposed adaptation network}
\label{fig:adaptation}
\vspace{-0.4cm}
\end{figure}

Figure\,\ref{fig:adaptation} illustrates an adaptation network where $C_{i}^{M_i}$ is the $i$-th convolutional layer of the network with a kernel size and stride of (1,1)
, and $M_i$ is the number of feature maps of the $i$-th convolutional layer. 

There are 48 possible patterns for each pixel in the proposed encryption, so each pixel has to be adjusted before DNNs are used.
$C_{i}^{M_i}$ learns the patterns of each encrypted pixel, and the feature representations of each encrypted pixel are then obtained.
As a result, the output of the adaptation network is applicable to any DNN.

\vspace{-0.2cm}
\section{Experiments and results}
\label{sec:experiments}

\subsection{Experimental Set-up}
\label{ssec:dataset}
To confirm that the proposed scheme is effective, we evaluated the performance in terms of image classification accuracy and compared it with conventional privacy-preserving methods.

We employed CIFAR10, which contains $32 \times 32$ pixel color images and consists of 
50,000 training images and 10,000 test images in 10 classes\cite{Krizhevsky09learningmultiple}.
For preprocessing, standard data augmentation (shifting and random horizontal flip) was used. 

The network was trained by using stochastic gradient descent (SGD) with momentum for 300 epochs.
The learning rate was initially set to 0.1 and was decreased by a factor of 10 at 150 and 225 epochs.
We used a weight decay of 0.0005, a momentum of 0.9, and a batch size of 128.

\begin{table}
\small
\centering
\caption{Image classification accuracy where data augmentation} was carried out in cloud server. (ResNet-18)
\label{tbl:result1}
\begin{tabular}{|c|c|}
\hline
Encryption & Accuracy (\%) \\
\hline
Plain Image & 92.98\\
\hline
Proposed (step 2) & \multirow{2}{*}{85.15}\\
without Adaptation&\\
\hline
Proposed (step 2) & \multirow{2}{*}{\textbf{86.99}}\\
with Adaptation&\\
\hline
Proposed (step 2 and 3) & \multirow{2}{*}{82.51}\\
without Adaptation&\\
\hline
Proposed (step 2 and 3) & \multirow{2}{*}{86.16}\\
with Adaptation&\\
\hline
Tanaka's Scheme\cite{tanaka}&56.41\\
\hline
EtC\cite{KURIHARA2015encryption,Kuri_2017}&69.03\\
\hline
Pixel-based\cite{2016_Gaata}&58.59\\
\hline
\end{tabular}
\vspace{-0.3cm}
\end{table}
\vspace{-0.3cm}
\subsection{Experimental Results}
\label{ssec:result}
The proposed encryption was used to encrypt all training and testing images, 
and networks were then trained and tested by using the encrypted images, as shown in Fig.\,\ref{fig:scenario}. In the experiment, the numbers of feature maps,
$M_1$ and $M_2$, were set to 8 and 32, respectively, and we evaluated the image classification accuracy of encrypted images 
under the use of ResNet-18\cite{resnet,zhang2018mixup}, which consists of 18 layers.
\vspace{-0.2cm}
\subsubsection{Data Augmentation in Cloud Server}
\vspace{-0.2cm}
Table\,\ref{tbl:result1} shows the classification accuracy in the case that the data augmentation was carried out in cloud server. 
Although the proposed scheme had a lower classification accuracy than that for non-encrypted images, it still maintained a high classification performance compared with 
the conventional methods.
Moreover, the use of the adaptation network was confirmed to improve the classification accuracy for two types of encryption: (Step 2) and (Step 2 and 3).
\vspace{-0.2cm}
\subsubsection{Data Augmentation in Client or Cloud Server}
\vspace{-0.2cm}
Table\,\ref{tbl:result2} shows that the accuracy of the proposed scheme was higher than that of Tanaka's scheme even when the data augmentation was carried out in a client.
In addition, the proposed scheme was confirmed to maintain a high classification performance, although Tanaka's scheme was heavily affected by the data augmentation.
\vspace{-0.2cm}
\subsubsection{Effects of Adaptation Network}
\vspace{-0.2cm}
When the data augmentation was applied to the images encrypted by using the conventional encryption\cite{tanaka}, the kernel of the block adaptation layer overlapped with the adjacent encrypted blocks.
As a result, the classification accuracy of the conventional encryption was heavily degraded. In comparison, the proposed encryption 
resulted in superior robustness against augmentation due to the use of pixel-based encryption and a kernel size of (1,1).

\begin{table}
\small
\centering
\caption{Image classification accuracy where data augmentation} was carried out in server or client. (ResNet-18)
\label{tbl:result2}
\begin{tabular}{|c|>{\centering\arraybackslash}m{2cm}|>{\centering\arraybackslash}m{2cm}|}
\hline
\multirow{2}{*}{Encryption} & \multicolumn{2}{c|} {Data Augmentation}\\
\hhline{~--}
&In Client&In Cloud\\
\hline
Proposed (Step 2)&\multirow{2}{*}{\textbf{92.94}}&\multirow{2}{*}{85.15}\\
without Adaptation&&\\
\hline
Proposed (Step 2)&\multirow{2}{*}{90.65}&\multirow{2}{*}{\textbf{86.99}}\\
with Adaptation&&\\
\hline
Tanaka's Scheme\cite{tanaka} &85.84&56.41\\
\hline
\end{tabular}
\vspace{-0.3cm}
\end{table}
\vspace{-0.3cm}
\section{Conclusion}
\label{sec:conclusion}
We presented a novel privacy preserving scheme for DNNs that enables us not only to use encrypted images on DNNs but to also use data augmentation in the encrypted domain.
This paper was the first to discuss data augmentation in the encrypted domain. Novel pixel-based image encryption was
proposed to protect the visual information of images and is available for training and testing DNNs. In addition, we proposed a novel adaptation network that 
enhances the classification performance of DNNs by obtaining representations of each pixel before passing through state-of-the-art DNNs. 
The experimental results showed that the proposed scheme performed better in classification than did the conventional block-based and pixel-based encryption schemes. In addition, 
it was proved that the proposed encryption allows us to use data augmentation in the encrypted domain. 
As a result, the advantages of data augmentation for enhancing the performance of DNNs can be efficiently utilized 
even when visual information is protected.

\begin{small}

\end{small}
\end{document}